\documentclass[prb,twocolumn,longbibliography,superscriptaddress]{revtex4-1}
\usepackage[utf8]{inputenc}
\usepackage{amsmath}
\usepackage{amsfonts}
\usepackage{amssymb}
\usepackage{bm}
\usepackage{graphicx}
\usepackage{xcolor}
\usepackage{microtype}
\usepackage{ dsfont }

\begin{document}
\title{Topological Mott transition in a Weyl-Hubbard model \\with dynamical mean-field theory}
\author{Bernhard Irsigler}
\affiliation{Institut f\"ur Theoretische Physik, Goethe-Universit\"at, 60438 Frankfurt am Main, Germany}
\author{Tobias Grass}
\affiliation{ICFO-Institut de Ciencies Fotoniques, The Barcelona Institute of Science and Technology, 08860 Castelldefels (Barcelona), Spain}
\author{Jun-Hui Zheng}
\affiliation{Center for Quantum Spintronics, Department of Physics,
Norwegian University of Science and Technology, NO-7491 Trondheim, Norway}
\author{Mathieu Barbier}
\affiliation{Institut f\"ur Theoretische Physik, Goethe-Universit\"at, 60438 Frankfurt am Main, Germany}
\author{Walter Hofstetter}
\affiliation{Institut f\"ur Theoretische Physik, Goethe-Universit\"at, 60438 Frankfurt am Main, Germany}
\begin{abstract}
Weyl semimetals are three-dimensional, topologically protected, gapless phases which show exotic phenomena such as Fermi arc surface states or negative magnetoresistance. It is an open question whether interparticle interactions can turn the topological semimetal into a topologically nontrivial Mott insulating phase. We investigate an experimentally motivated model for Weyl physics of cold atoms in optical lattices, with the main focus on interaction effects and topological properties by means of dynamical mean-field theory (DMFT). We characterize topological phases by numerically evaluating the Chern number via the Ishsikawa-Matsuyama formula for interacting phases. Within our studies, we find that the Chern numbers become trivial when interactions lead to insulating behavior. For a deeper understanding of the  Weyl-semimetal-to-Mott-insulator topological phase transition, we evaluate the topological properties of quasiparticle  bands as well as so-called blind bands. Our study is complementary to recent studies of Weyl semimetals with DMFT.
\end{abstract}
\maketitle

\section{Introduction}
Topological states of matter realized with cold atoms in optical lattices are a vibrant field at the forefront of modern quantum research 
\cite{Dalibard2011,Goldman2014,Aidelsburger2018,Hofstetter2018,Cooper2019}. The great control and tunability of cold atoms in optical lattices make them an ideal analog quantum simulator of tight-binding Hamiltonians 
\cite{Bloch2005,Bloch2008}. Among pioneering experiments in the context of topological states are the realizations of two prominent theoretical two-dimensional (2d) models: the Hofstadter 
\cite{Hofstadter1976} and the Haldane model 
\cite{Haldane1988}. The former is realized by imprinting a complex quantum phase onto the particles upon hopping in the lattice through laser-assisted tunneling 
\cite{Aidelsburger2013,Miyake2013}. The latter is engineered through elliptic lattice shaking which also imprints a complex phase according to Floquet's theorem 
\cite{Jotzu2014,Flaschner2016}. Both approaches are well described by effective static Floquet Hamiltonians with gauge fields as a result of high-frequency driving 
\cite{Bukov2015a,Eckardt2017}.

The current focus of research in this field clearly lies in 2d systems. One reason is the fact that 2d systems host paradigmatic phases such as the quantum Hall effect. The possible existence of topological phases is connected to the dimensionality and symmetries of the system of interest 
\cite{Ryu2010}. In contrast to 2d, in three-dimensional (3d) systems, even gapless states can be topologically protected. Examples are the Weyl semimetal (WSM) and nodal-line semimetals
\cite{Feng2016,Armitage2018}. Moreover, the search for an exotic topological Mott insulator suggested its existence in 3d only\cite{Pesin2010,Rachel2018}.

WSMs host gapless Weyl points (WPs) in the 3d Brillouin zone (BZ) which are topologically protected, i.e., they cannot gap out through smooth deformations of the Hamiltonian. One generally differentiates between WSMs with broken time-reversal symmetry or WSMs with broken inversion symmetry
\cite{Armitage2018}. If both are broken, the WPs are not located at the Fermi level
\cite{Zyuzin2012}. WSMs have first been observed in 2015 in a TaAs crystal along with the exotic Fermi arc surface states by means of photoemission spectroscopy
\cite{Xu2015} as well as in a gyroid photonic crystal
\cite{Lu2015} both with broken inversion symmetry. Another intriguing feature of WSMs is the chiral anomaly and the resulting negative magnetoresistance which was also measured in TaAs crystals
\cite{Zhang2016a}. Recently, a nodal-line semimetal has been engineered as the first instance of a 3d topological state in a cold atom setup\cite{Song2019}, but the realization of an atomic WSM is still lacking.

In the interacting case, the Weyl-Mott insulator has been proposed as an extension to the noninteracting WSM
\cite{Morimoto2016}.
The model has a momentum-locked interaction and is analytically solvable. This is possible through the assumption of this particular form of the interaction. Moreover, the system has a Mott gap as well as a nontrivial topological invariant in terms of the single particle Green's function.
Ref.~
\onlinecite{Yang2019} pointed out that this invariant does not imply the presence of a single-particle Fermi arc because of the absence of the WPs in the single-particle spectrum. Instead, the system has gapless particle-hole pair excitations, suggesting the existence of the Weyl points in the bosonic excitation spectrum. The nonzero topological invariant indeed implies the presence of a {bosonic surface state}. While a single-particle Fermi arc is observable through photoemission spectroscopy
\cite{Xu2015}, the bosonic surface is not accessible with photoemission spectroscopy.

In Ref.~
\onlinecite{Morimoto2016}, the interactions which give rise to the Weyl-Mott insulator are local in momentum space, whereas in realistic systems, the interactions are  rather local in real space. In the present paper, we investigate the effect of realistic on-site interactions on a WSM. To analyze the topological properties of such a system, we compute the topological invariants in terms of the single-particle Green's function. In most cases, this quantity is well suited to examine the topologically non-trivial behavior. This evaluation is particularly useful when the many-body wavefunction is numerically not accessible.
We use dynamical mean-field theory (DMFT) in order to solve the present many-body problem approximately
\cite{Georges1996}. In the context of topological systems, DMFT has been used in numerous studies in 2d
\cite{Cocks2012,Orth2013,Vasic2015,Amaricci2015,Vanhala2016,Kumar2016,Amaricci2017,Zheng2018,
Irsigler2019,Irsigler2019a} as well as 3d systems
\cite{Amaricci2016,Irsigler2020}.  DMFT has been applied recently to WSMs: In Ref.~\onlinecite{Crippa2020}, the nonlocal annihilation of WPs within the BZ has been observed which is impossible in the noninteracting case. Ref.~
\onlinecite{Acheche2020}, on the other hand, investigated the influence of interactions in view of the negative magnetoresistance. Our focus lies on the topological properties of the  many-body phases which we obtain. We find that the WSM is robust up to a critical interaction strength. In particular, we observe that the transition from a topologically nontrivial WSM to a trivial Mott insulator occurs through the emergence of pairs of quasiparticle bands and so-called blind bands. Here, the former are topologically nontrivial and cancel out the nontrivial properties of the original WSM while the latter are topologically trivial. This ultimately results in an overall topologically trivial Mott insulator.

The article is structured as follows: In Sec.~\ref{mod}, we introduce the model for a WSM and investigate its noninteracting properties. In Sec.~\ref{mott}, we analyze the WSM-to-Mott-insulator transition of the interacting model. In Sec.~\ref{IM}, we compute topological properties as a function of the interaction strength. In Sec.~\ref{quasi}, we discuss the effective quasiparticle spectrum and elaborate on the interaction-induced WSM-to-Mott-insulator topological phase transition. Finally, we conclude in Sec.~\ref{con}.

\section{Model}
\label{mod}
We study the tight-binding model proposed by Dub\v{c}ek et al. 
\cite{Dubcek2015} which is motivated by the experimental implementation of the Hofstadter model in Ref.~
\onlinecite{Miyake2013}, extended to three spatial dimensions. The corresponding real-space Hamiltonian reads
\begin{equation}
\begin{split}
\hat{H}_\text{Dubcek}=-\sum_{\bm{j}}&\left[(-1)^{x+y}K_x\hat{c}_{\bm{j}+\hat{\bm{x}}}^\dag\hat{c}_{\bm{j}}
+J_y\hat{c}_{\bm{j}+\hat{\bm{y}}}^\dag\hat{c}_{\bm{j}}\right.\\
&\left.+(-1)^{x+y}K_z\hat{c}_{\bm{j}+\hat{\bm{z}}}^\dag\hat{c}_{\bm{j}}+\text{h.c.}\right]
\end{split}
\label{DubcekHamReal}
\end{equation}
where $\bm{j}=(x,y,z)$ is a 3d lattice vector on a cubic lattice, $\hat{c}_{\bm{j}}$ ($\hat{c}_{\bm{j}}^\dag$) annihilates (creates) a fermion at lattice site $\bm{j}$, and $\hat{\bm{\nu}}$ denotes the unit vector in $\nu$~direction. In the following, we focus on the isotropic case and set the hopping energies to the unit of energy $K_x=J_y=K_z=1$. The momentum-space Hamiltonian reads
\begin{equation}
H_\text{Dubcek}(\bm{k})=-2\left[\cos(k_y)\sigma^x+\sin(k_x)\sigma^y-\cos(k_z)\sigma^z\right]
\label{DubcekHam}
\end{equation}
where we have set the lattice constant to unity. Here, the Pauli matrices $\sigma^\nu$ refer to the pseudo-spin space of the two sites of the unit cell which breaks inversion symmetry. We read off four degeneracies of the Hamiltonian in Eq.~\eqref{DubcekHam} at  the points $(k_x,k_y,k_z)=(0,\pm\pi/2,\pm\pi/2)$ in the first BZ. To confirm whether these degeneracies are indeed WPs, we compute the Chern number on a closed surface around a single degeneracy using Fukui's method 
\cite{Fukui2005}. In fact, any smooth closed surface can be used, see appendix~\ref{appA}. Indeed, the four points $(0,\pm\pi/2,\pm\pi/2)$ exhibit nonzero Chern numbers (+1 or -1), also dubbed \textit{topological charge}. The sum over the four topological charges is zero.

\section{Mott transition}
\label{mott}
Let us now focus on the properties of the Mott transition of the model in Eq.~\eqref{DubcekHamReal}. We consider fermions with a Hubbard interaction term $U\sum_{\bm{j}}\hat{n}_{\bm{j}\uparrow}\hat{n}_{\bm{j}\downarrow}$ where $U$ is the interaction strength and $\hat{n}_{\bm{j}\sigma}=\hat{c}_{\bm{j}\sigma}^\dag\hat{c}_{\bm{j}\sigma}$ is the particle number operator of a spin-$\sigma$ fermion on lattice site $\bm{j}$.  Spin states are introduced in the following way in the four-band interacting Hamiltonian:
\begin{equation}
\hat{H}_\text{int}=\begin{pmatrix}
\hat{H}_\text{Dubcek}&0\\
0&\hat{H}_\text{Dubcek}\\
\end{pmatrix}
+U\sum_{\bm{j}}\hat{n}_{\bm{j}\uparrow}\hat{n}_{\bm{j}\downarrow}
\label{interactingHam}
\end{equation}
The spin degeneracy results in a factor of 2 for the topological charges of the WPs. 

One of the most successful methods for investigating Hubbard-like Hamiltonians and describing their Mott transitions is DMFT
\cite{Georges1996}. It maps the full Hubbard model onto a set of coupled self-consistent quantum impurity models which can be solved through different approaches like quantum Monte Carlo
\cite{Gull2011} or exact diagonalization
\cite{Caffarel1994} (ED). This mapping neglects  nonlocal fluctuations but keeps track of all local quantum fluctuations. This manifests in a momentum-independent selfenergy $\Sigma^{\sigma\sigma'}(\omega,\bm{k})=\Sigma^{\sigma\sigma'}(\omega)$ with $\sigma$ and $\sigma'$ denoting spin states. As in static mean-field theories, DMFT is solved self-consistently and thus depends on an initial guess.

Here, we perform real-space DMFT
\cite{Helmes2008,Snoek2008} calculations on a $6\times6\times6$ lattice for the model in Eq.~\eqref{DubcekHam} with an ED solver with four bath sites. We are interested in the paramagnetic case.  The paramagnetic solution is sufficient to describe the Mott transition. Besides, the temperature regimes we consider are above the superexchange temperature for antiferromagnetic ordering. The paramagnetic solution is found if diagonal elements of the selfenergy in spin space are identical and off-diagonal elements vanish:
\begin{align}
\Sigma^{\uparrow\uparrow}(\omega)&=\Sigma^{\downarrow\downarrow}(\omega)\equiv\Sigma(\omega)\\\Sigma^{\uparrow\downarrow}(\omega)&=\Sigma^{\downarrow\uparrow}(\omega)=0
\end{align}

The Hamiltonian in Eq.~\eqref{interactingHam} is symmetric under the translations $\bm{j}\rightarrow\bm{j}+\hat{\bm{z}}$ and $\bm{j}\rightarrow\bm{j}+\hat{\bm{x}}+\hat{\bm{y}}$. It is then sufficient to compute only two separate local selfenergies, i.e., solving two separate impurity problems, and copy them accordingly in the lattice Green's function.
\begin{figure}
\centering
\includegraphics[width=\columnwidth]{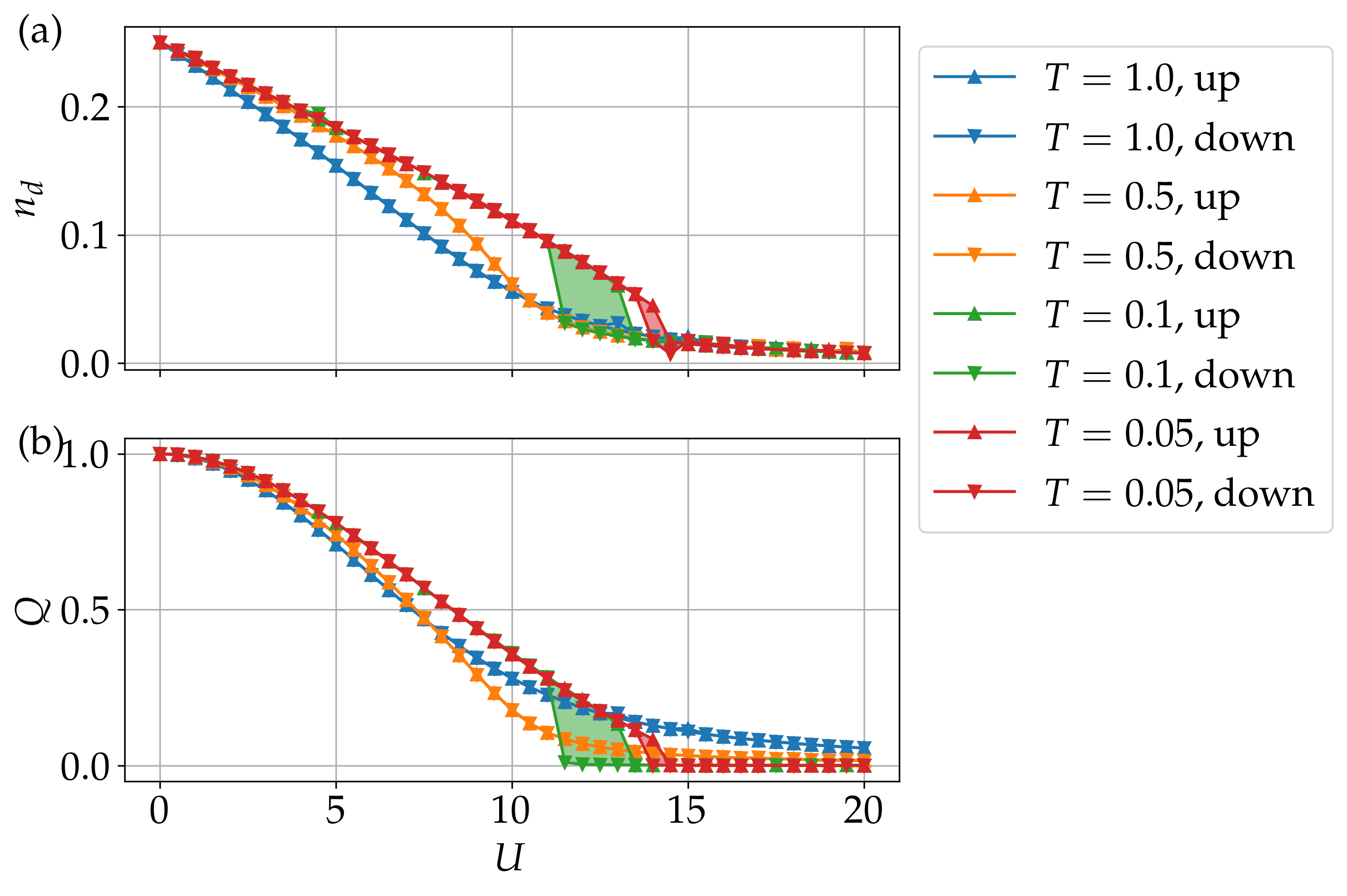}
\caption{Double occupancy $n_d$ in (a) and quasiparticle weight $Q$ in (b) as functions of the interaction strength $U$ for different temperatures $T$. The direction of consecutive initial guesses for the selfenergy for the DMFT calculations is labeled by \textit{up} and \textit{down}. The hysteresis between those is highlighted by a shaded area. Energies are measured in units of the hopping energy.}
\label{Fig:doubleOccupancy}
\end{figure}

As indicators for the Mott transition, we compute two quantities:  (i) the double occupancy 
\begin{equation}
n_d=\frac{1}{N_s}\sum_{\bm{j}}\langle\hat{n}_{\bm{j}\uparrow}\hat{n}_{\bm{j}\downarrow}\rangle
\end{equation} 
where $N_s$ is the number of lattice sites and $\langle\dots\rangle$ denotes the ensemble average; (ii) the quasiparticle weight~\cite{Georges1996}, defined as
\begin{equation}
Q=\left[1-\left.\frac{\partial\Sigma(\omega)}{\partial\omega}\right|_{\omega=0}\right]^{-1}
=\left[1-\left.\frac{\Sigma(i\omega_n)}{i\omega_n}\right|_{n=0}\right]^{-1}
\end{equation}
where we have introduced the real-frequency selfenergy $\Sigma(\omega)$ and the selfenergy in terms of Matsubara frequencies $\Sigma(i\omega_n)$. We present the results for  $n_d$ and $Q$ in Fig.~\ref{Fig:doubleOccupancy} as functions of the interaction strength $U$ for different temperatures. The self-consistent solutions are found successively for different $U$. The initial guess for the self-consistent DMFT iteration is inherited from the previous converged solution for the previous value of $U$. Starting with $U=0$, i.e., going upwards, the first guess for the initial selfenergy is zero. For the downwards calculations, the deep Mott solution at $U=20$ was used which was previously found by the upwards calculation. As the difference between those curves, we observe the typical hysteresis of the paramagnetic solutions shown as shaded areas
\cite{Georges1996}. The hysteresis reflects the coexistence of two solutions, i.e., the correlated WSM and the Mott insulator. The critical interaction strength for this phase transition is located within this coexistence regime. As we observe in Fig.~\ref{Fig:doubleOccupancy}, this regime is temperature dependent, and thus also the critical interaction strength. For comparison, the critical interaction strength for the metal-to-Mott-insulator transition in the 3d Hubbard model at $T\approx0.33$ is $U=15.4$
\cite{Lichtenstein2004}.

\begin{figure}
\centering
\includegraphics[width=.8\columnwidth]{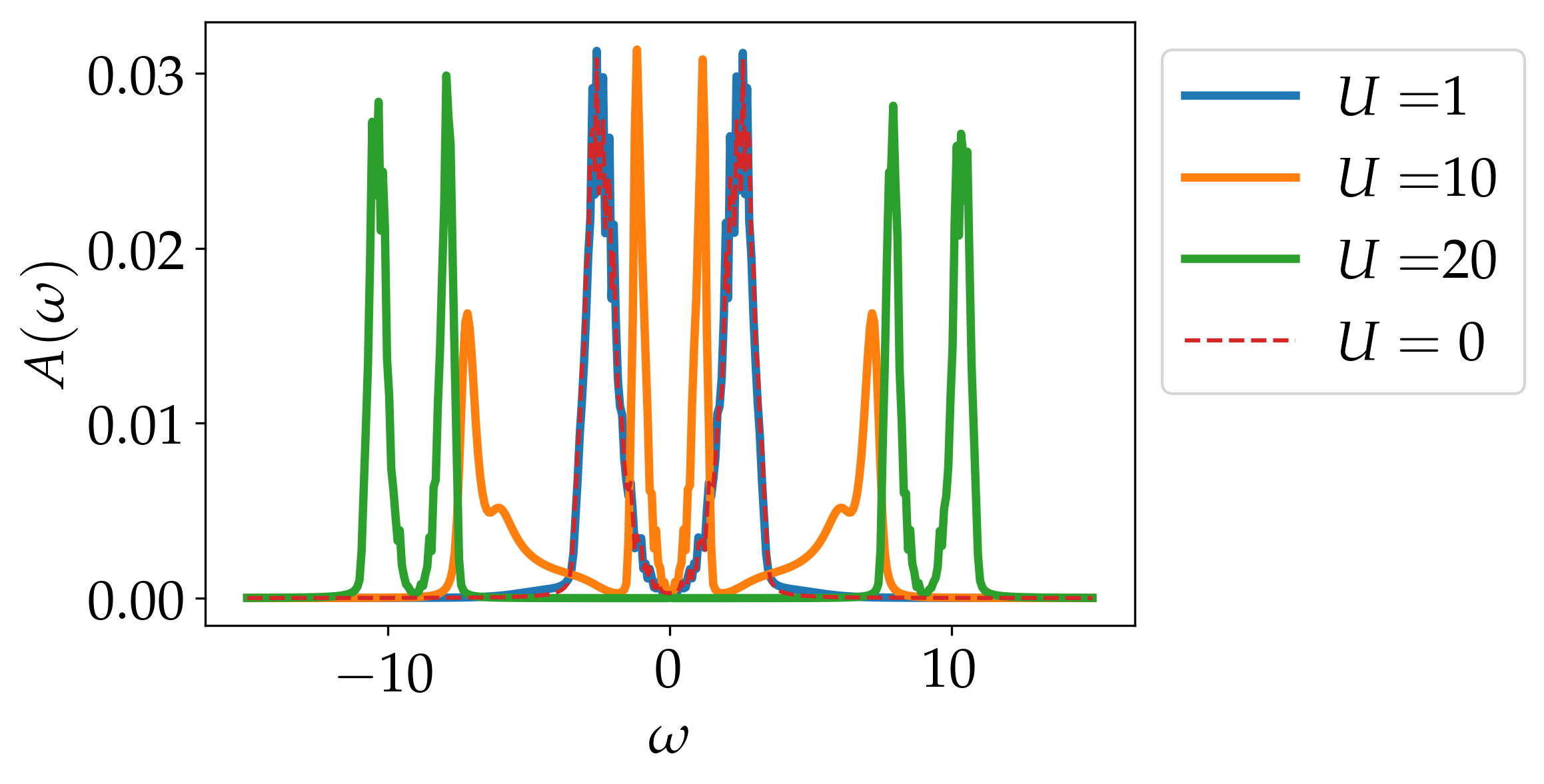}
\caption{Density of states $A(\omega)$ as a function of the frequency $\omega$ for different $U$ at $T=0.1$.}
\label{Fig:densityOfStates}
\end{figure}
To determine transport properties of the obtained many-body phases, we are interested in the  density of states
\begin{equation}
A(\omega)=-\frac{1}{\pi}\text{Im}\int d\bm{k}\text{Tr}G(\omega,\bm{k})
\label{DOS}
\end{equation}
where we have defined the retarded, real-frequency single-particle Green's function
\begin{equation}
G(\omega,\bm{k})=\frac{1}{[\omega+0^+-\Sigma(\omega)+\mu]\mathds{1}-{H}(\bm{k})},
\label{GF}
\end{equation}
which does only apply for the paramagnetic solutions. Here, $\mathds{1}$ denotes the $2\times2$ identity matrix in the sublattice representation, and $\mu$ is the chemical potential which is set to $U/2$ throughout the article, constraining the system to be half-filled. 
In Fig.~\ref{Fig:densityOfStates}, we show the density of states for different~$U$ at $T=0.1$. For $U=1$, the density of states is almost identical to the one of the noninteracting case $U=0$. This is expected since the selfenergy is small in this regime. We also observe the peaks from the two bands of the Hamitonian and an approximately quadratic behavior around $\omega=0$ which is a property of a semimetal. For $U=10$, we observe two Hubbard bands at approximately $\omega=\pm8$. The bands close to $\omega=0$ are shrunk compared to the $U=1$ case but the system is still semimetallic. For $U=20$, we find an overall gap of size $\sim 16$ which corresponds to the Mott gap. The structure of each of the Hubbard bands resembles the structure of the original density of states at $U=0$. Such splitting of the noninteracting bands, each with the density of states similar to the noninteracting one, has been observed before in a bosonic system
\cite{Vasic2015}.

In Eq.~\eqref{GF}, the selfenergy in terms of real frequencies $\omega$ enters. Most impurity solvers, however, provide the output as a function of Matsubara frequencies $i\omega_n$. Here, we use the maximum entropy method
\cite{Jarrell1996} in order to map $\Sigma(i\omega_n)$ to $\Sigma(\omega)$. This method was originally developed to analytically continue noisy quantum Monte Carlo data. It has the advantage to yield smooth outcomes through Bayesian statistics. Here, we use this method to analytically continue ED results. Conventionally, the density of states from ED calculations is rugged due to the finite number of bath sites. Here, the maximum entropy method can compensate that. Of course, the result is then approximate. The results in Fig.~\ref{Fig:densityOfStates} show that our approach of combining the maximum entropy method with ED results yields a reasonable outcome. 

In summary, the double occupancy, the quasiparticle weight, and the density of states provide clear evidence that the many-body phase for strong $U>15$ is a Mott insulator. Let us now turn to the topological properties of the interacting system.

\section{Ishikawa-Matsuyama formula}
\label{IM}

The Ishikawa-Matsuyama formula manifests the generalization of a Chern number to interacting systems as it corresponds to the Hall conductivity up to a constant factor and is formulated in terms of Green's functions 
\cite{Ishikawa1986}:
\begin{equation}
C_\text{IM}= \frac{\epsilon^{\nu\rho\eta}}{24\pi^2}\int dk\text{Tr}\left\{\left[{G}\partial_\nu {G}^{-1}\right]\left[{G}\partial_\rho {G}^{-1}\right]\left[{G}\partial_\eta {G}^{-1}\right]\right\}
\label{IMformula}
\end{equation}
where $k=(k_0,k_1,k_2)$ with $k_0=i\omega_n$ and $\nu,\rho,\eta$ run over the elements of $k$. We also have used the abbreviation $G=G(k)=G(i\omega_n,k_1,k_2)$. The formula is rather complicated compared to the noninteracting TKNN invariant
\cite{Thouless1982}. It has been shown, however, that  in some regimes the information about the full frequency range is not necessary and only the $\omega=0$ mode is crucial
\cite{Wang2012}. This is called the effective topological Hamiltonian approach which makes it possible to compute topological invariants from an effective, noninteracting Hamiltonian \mbox{$H_\text{top}=-G^{-1}(\omega=0,\bm{k})$}. This, however, is valid only if the Green's function has no zeros which is of course not the case in a Mott insulator. Thus we have to consult the formula in Eq.~\eqref{IMformula}.
To this end, we define  the single-particle Green's function  within the DMFT framework, i.e., $\Sigma(i\omega_n,\bm{k})=\Sigma(i\omega_n)$, according to Ref.
\onlinecite{Zheng2019}
\begin{equation}
{G}(i\omega_n,k_1,k_2)=\frac{1}{[i\omega_n-{\Sigma}(i\omega_n)+\mu]\mathds{1}-{H}(k_1,k_2)}
\label{matsubaraGF}
\end{equation}
For the sake of brevity, we drop all the arguments. So, we find
\begin{align}
\partial_{k_0}{G}^{-1}&=(1-\partial_{k_0}{\Sigma})\mathds{1}=(1+i\partial_{\omega_n}{\Sigma})\mathds{1}\\
\partial_{k_\nu}{G}^{-1}&=\partial_{k_\nu}{H}={j}_\nu
\end{align}
where ${j}_\nu=j_\nu(k_1,k_2)$ is the current in $\nu=1,2$ direction with the  2d momenta  $k_1$ and $k_2$.  Consequently, Eq.~\eqref{IMformula} simplifies to 
\begin{equation}
\begin{split}
C_\text{IM}=\frac{i}{8\pi^2}\int &dk_1dk_2d\omega_n \\
\times\text{Tr}&\left[{G}{j}_1{G}{j}_2{G}(1+i\partial_{\omega_n}{\Sigma})\right.\\
&\left.-(1+i\partial_{\omega_n}{\Sigma}){G}{j}_2{G}{j}_1{G}\right]
\end{split}
\label{IManalytical}
\end{equation}

Following the above discussion of the noninteracting case, see also appendix~\ref{appA}, we will put the 2d momentum $(k_1,k_2)$ onto a surface enclosing the WPs in the 3d BZ of the interacting system to compute topological charges of the WPs in the interacting case.

The momentum-dependent part of the formula in Eq.~\eqref{IManalytical} can be calculated analytically depending on the surface  enclosing the WP over which we want to integrate. For the two components of the currents, this implies
\begin{equation}
j_r =  \sum_{\nu}j_\nu\frac{\partial k_\nu}{\partial k_r}, \quad r=1,2\text{ and } \nu=x,y,z.
\end{equation}
The frequency derivative is performed numerically as
\begin{equation}
\partial_{\omega_n}\Sigma(i\omega_n)\approx\frac{\Sigma(i\omega_{n+1})-\Sigma(i\omega_n)}{2\pi T},
\end{equation}
according to the definition of the fermionic Matsubara frequencies $\omega_n=\pi T(2n+1)$ with $n$ being an integer. Before computing the topological charge of the interacting system by enclosing the WPs with a surface, we have to find their position within the BZ as a function of $U$ because their position could, in general, dependent on the interaction. To this end, we maximize the imaginary part of the Green's function at the Fermi level $-\text{Im}\text{Tr}G(\omega=0,\bm{k})$ which corresponds to the contribution to the density of states, see Eq.~\eqref{DOS}. The obtained momentum yields the position of the WPs. Interestingly, as the result, we find that the  position of the WPs does not depend on the strength of the interaction, which is not shown here. However, we note that the inclusion of a staggered potential as, e.g., in Ref.~\onlinecite{Kumar2016}, might change this since it is another energy scale competing with the interaction strength. Also note that the described procedure of determining the positions of the WPs does not rely on an effective noninteracting theory, in contrast to the procedure of
Ref.~\onlinecite{Crippa2020}.

\begin{figure}
\centering
\includegraphics[width=\columnwidth]{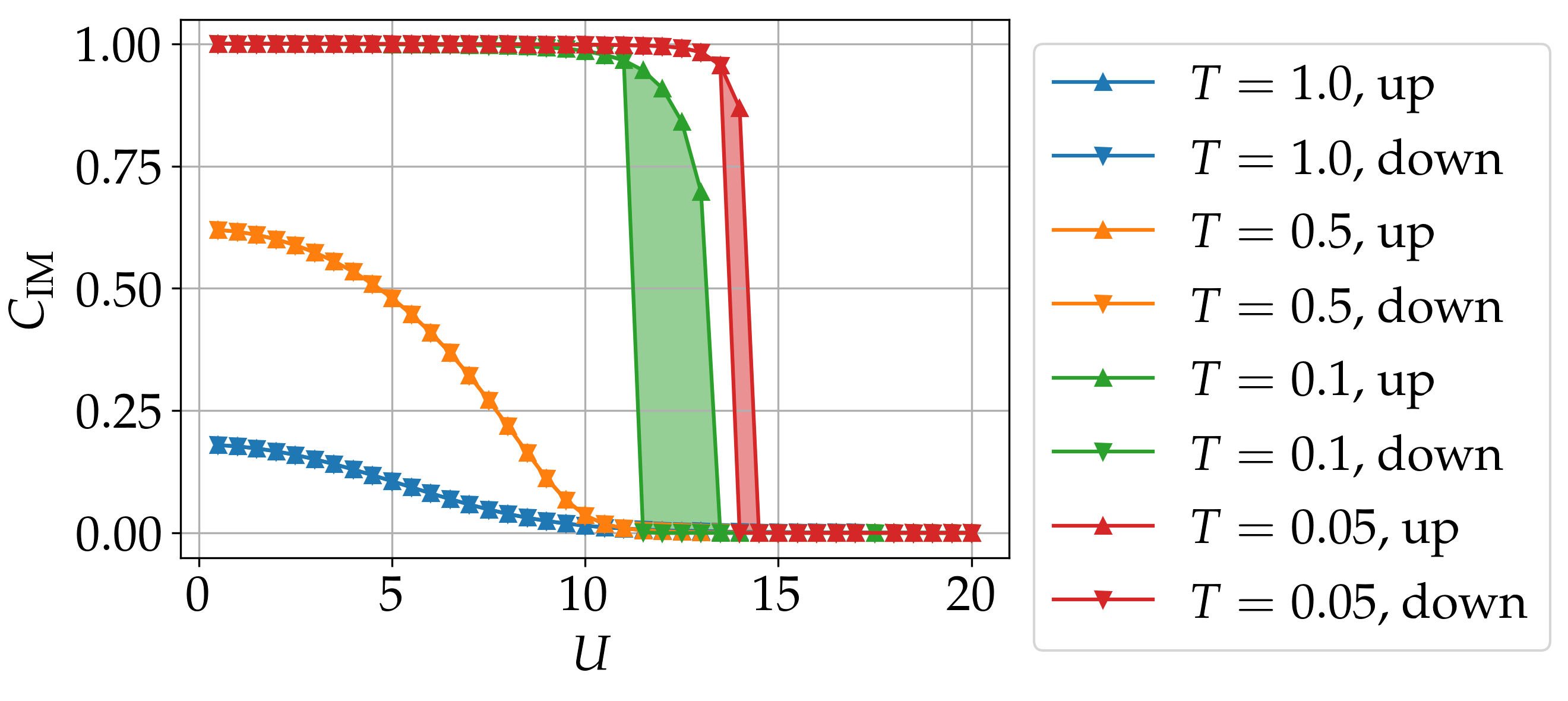}
\caption{Ishikawa-Matsuyama invariant $C_\text{IM}$ for the interacting Hamiltonian in Eq.~\eqref{interactingHam} as a function of $U$ for different temperatures on a sphere in the BZ with radius $R_S=\pi/2$ enclosing the WP at $\bm{k}_\text{WP}=(0,\pi/2,\pi/2)$ for one spin state. Here, we again highlight the hysteresis by a shaded area.}
\label{Fig:invariants}
\end{figure}

In Fig.~\ref{Fig:invariants}, we show the Ishikawa-Matsuyama invariant $C_\text{IM}$ calculated on a sphere with radius $R_S=\pi/2$ enclosing the WP at $\bm{k}_\text{WP}=(0,\pi/2,\pi/2)$. For moderate as well as large $U$, we find well quantized results. Close to the Mott transition, the $C_\text{IM}$ is not quantized anymore. This is  due to the finite temperature which becomes comparable to the gap in the vicinity of the phase transition, see Fig.~\ref{Fig:densityOfStates}.

\section{Quasiparticle spectrum and blind bands}
\label{quasi}
It is anticipated that the topological invariant on the enclosing surface vanishes when the WPs gap out since the system then lacks the singularity which has to be enclosed, compare Figs.~\ref{Fig:densityOfStates} and~\ref{Fig:invariants}. The resulting many-body state is globally gapped. Due to the lack of WPs there are neither sources nor sinks of Berry curvature. The many-body state is thus topologically trivial. For finite magnetization, topologically trivial
~\cite{Cocks2012,Kumar2016} as well as nontrivial~
\cite{He2011,Radic2012,Wu2016a,Gu2019,Ebrahimkhas2020} states have been found.

We want to understand in more detail how this topological phase transition to a topologically trivial Mott insulator occurs. To this end, we again focus on the paramagnetic case. 
Our conventional understanding of topological phase transitions is the closing of a quasiparticle band gap. Quasiparticle bands exhibit Chern numbers and correspond to the poles of the single-particle Green's function. It has been discussed, however, on the level of single-particle Green's functions, that not only poles of the Green's function can exhibit nontrivial Chern numbers but also zeros of the Green's function. The zeros of the Green's function are dubbed \textit{blind bands}. 
Ref.~\onlinecite{Gurarie2011} proposed the interaction-induced topological phase transition through a gap closing of blind bands. Herein, not only the quasiparticle bands, but also the blind bands exhibit nontrivial Chern numbers. The gap closing of blind bands then can induce a topological phase transition. In our case, we do not find nontrivial blind bands but rather a topological phase transition stemming from the quasiparticle bands only.

The topological properties of the interacting system are described by a formula for a generalized Chern number $\tilde{C}$ which relates the Chern numbers of quasiparticle  bands and the Chern numbers of blind bands and was derived from the Ishikawa-Matsuyama formula
\cite{Zheng2017}:
\begin{equation}
\begin{split}
\tilde{C} =& \sum_{n=1}^{N}\int dk_1dk_2\text{Im}\langle \partial_{k_1}\psi(\omega^p_n(\bm{k}),\bm{k})|\partial_{k_2}\psi(\omega^p_n(\bm{k}),\bm{k})\rangle\\
-&\sum_{m=1}^{M}\int dk_1dk_2\text{Im}\langle \partial_{k_1}\psi(\omega^z_m(\bm{k}),\bm{k})|\partial_{k_2}\psi(\omega^z_m(\bm{k}),\bm{k})\rangle
\end{split}
\label{JunHuiEq}
\end{equation}
Herein, we have defined the eigenstates $|\psi_j(\omega,\bm{k})\rangle$ of the Green's function according to
\begin{equation}
{G}(\omega,\bm{k})|\psi_j(\omega,\bm{k})\rangle
=g_j(\omega,\bm{k})|\psi_j(\omega,\bm{k})\rangle.
\end{equation}
Since the Green's function is not hermitian away from $\omega=0$, the eigenvalues $g_j(\omega,\bm{k})$ are not real in general and there is no generic ordering. Since we are only interested in zeros and poles of $g_j(\omega,\bm{k})$, we order the eigenvalues by their absolute values. In Eq.~\eqref{JunHuiEq}, we have also defined the quasiparticle  bands $\omega^p_n(\bm{k})$ and the blind bands $\omega^z_m(\bm{k})$ as the poles and zeros of the Green's function, respectively: 
\begin{equation}
g_j(\omega=\omega^p_n(\bm{k}),\bm{k})\rightarrow\infty
\text{ and }
g_j(\omega=\omega^z_m(\bm{k}),\bm{k})=0.
\end{equation}
We have dropped the band index $j$ for the states in Eq.~\eqref{JunHuiEq} since $j$ is fully determined by $\omega^p_n(\bm{k})$ and $\omega^z_m(\bm{k})$, respectively. Furthermore, we focus on the weakly interacting case and the deep Mott-insulating case. In the intermediate regime, the poles and zeros are not sufficiently pronounced. Note that the physics in the deep Mott regime will certainly differ from this treatment as, e.g., particle-hole excitations are neglected. We emphasize that our discussion  focuses on the framework of single-particle Green's functions.

\begin{figure}
\centering
\includegraphics[width=.9\columnwidth]{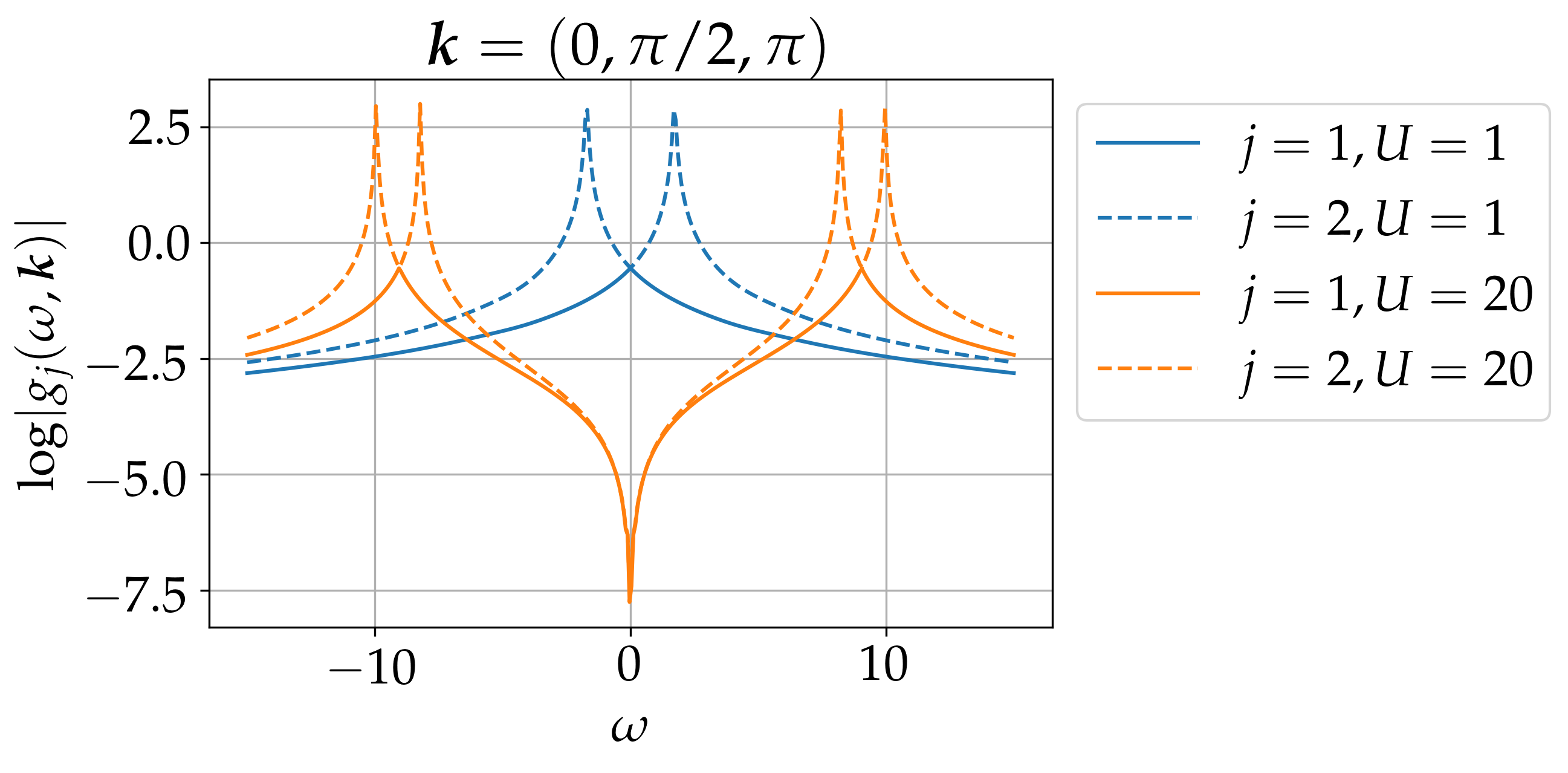}
\caption{Absolute value of the eigenvalues of the Green's function in log scale exemplarily for a specific $\bm{k}$ on the WP-enclosing sphere.}
\label{Fig:blindBands}
\end{figure}

We show the absolute value of the eigenvalues of the Green's function in Fig.~\ref{Fig:blindBands} as a function of $\omega$ exemplarily for $(k_1,k_2)=(0,0)$ on the WP-enclosing sphere which corresponds to $\bm{k}=(0,\pi/2,\pi)$, see appendix~\ref{appA} for details. For $U=1$, there are two poles corresponding to two quasiparticle  bands. These bands approximately correspond to the noninteracting energy bands since the interaction is small compared to the bandwidth. Poles in this plot are finite since we use a finite broadening factor $\eta$ in the analytically continued Green's function ${G}(\omega+i\eta,\bm{k})$ with the definition in Eq.~\eqref{matsubaraGF}. Also, the exact pole will not be matched perfectly because of the equidistant discretization of the frequency axis. For $U=20$, we observe four poles and additionally a zero at $\omega\approx0$. We also observe that the zero is doubly degenerate.

\begin{figure}
\centering
\includegraphics[width=.9\columnwidth]{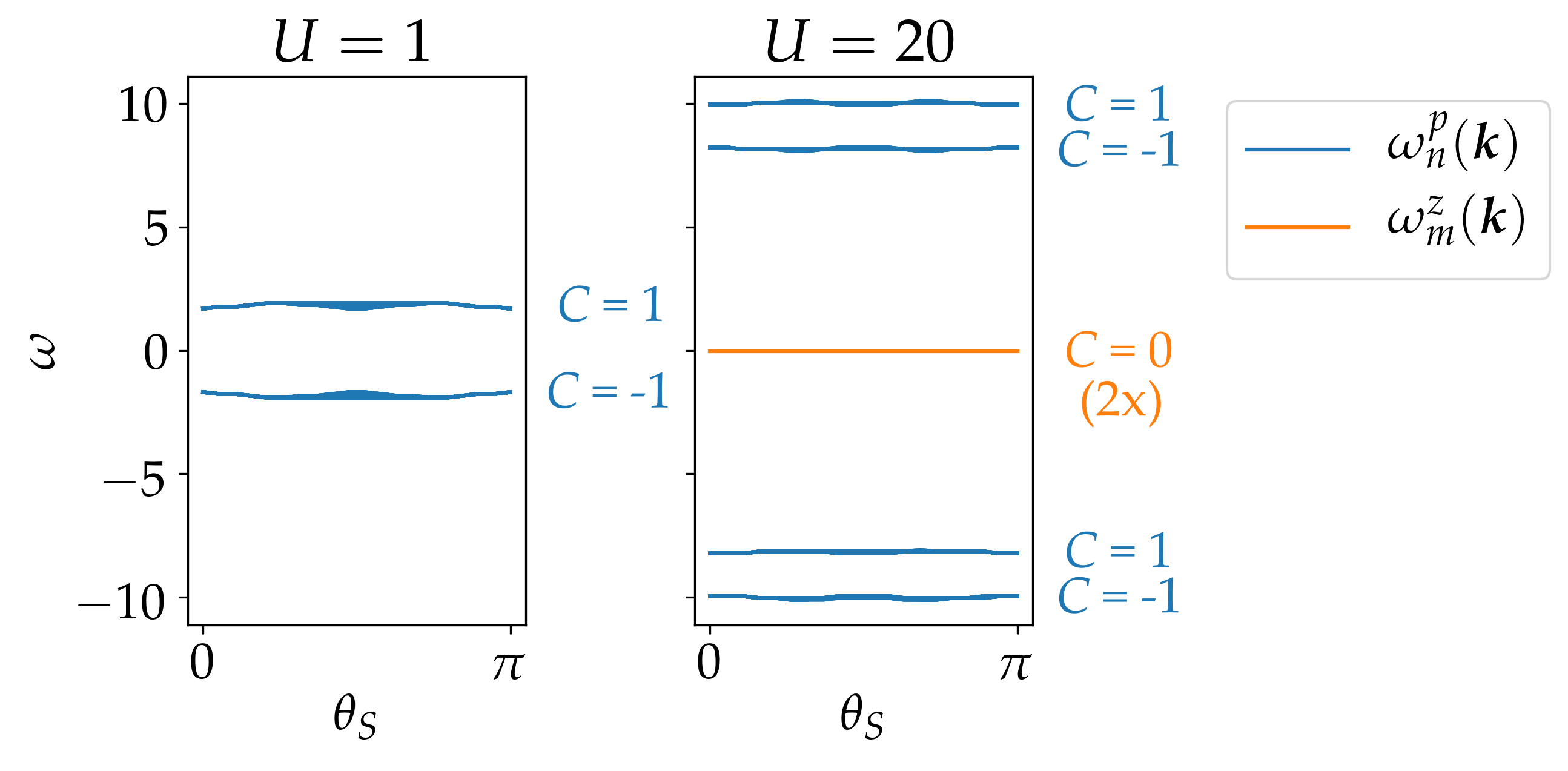}
\caption{Quasiparticle bands $\omega^p_n(\bm{k})$ and blind bands $\omega^z_m(\bm{k})$ of the single-particle Green's function on a sphere with radius $R_S=\pi/2$ enclosing the WP at $(0,\pi/2,\pi/2)$ in the BZ. The Chern number $C$ is written next to the respective band. Note that this 2d plot shows a function of the azimuth $k_1=\theta_S$ only. Values as a function of the polar angle are plotted implicitly on top of each other.}
\label{Fig:blindBandSpectrum}
\end{figure}

In Fig.~\ref{Fig:blindBandSpectrum}, we show the numerically determined momentum-resolved quasiparticle  bands $\omega^p_n(\bm{k})$ in blue and blind bands  $\omega^z_m(\bm{k})$ in orange of the single-particle Green's function. The respective Chern number $C$ is computed with the Fukui method \cite{Fukui2005} and is written next to the band. For $U=1$, the spectrum resembles that of the noninteracting case which is expected for such small interaction strength. Also, the Fermi level lies between the two bands which carry opposite nontrivial Chern numbers. This is consistent with a topologically nontrivial many-body phase, see Fig.~\ref{Fig:invariants}.

For $U=20$, we observe four quasiparticle  bands and a two-fold degenerate blind band. This shows the preserved difference $N-M$ between the number of quasiparticle  bands and the number of blind bands. We also note that the blind band is flat. This is because in the single-particle Green's function, Eq.~\eqref{matsubaraGF}, a zero emerges only if the selfenergy diverges. As the selfenergy is momentum-independent within DMFT, the blind band has no momentum dependence and is thus flat.
Additionally, the blind bands contribute zero Chern number to the total Chern number. Out of the four quasiparticle  bands, the lower two are occupied which have opposite Chern numbers. The total Chern number is thus zero which is consistent with the obtained topologically trivial Mott insulator, see Fig.~\ref{Fig:invariants}. Each of the Hubbard bands consists of subbands with  the same quasiparticle spectrum as the original noninteracting band structure. Since the sum of Chern numbers of all the bands in the original band structure is zero, the Hubbard bands are topologically trivial as well. This agrees with the topologically trivial Mott insulator found in the bosonic Haldane-Hubbard model studied with DMFT which showed the equivalent structure of subbands
\cite{Vasic2015}.

\begin{figure}
\centering
\includegraphics[width=.6\columnwidth]{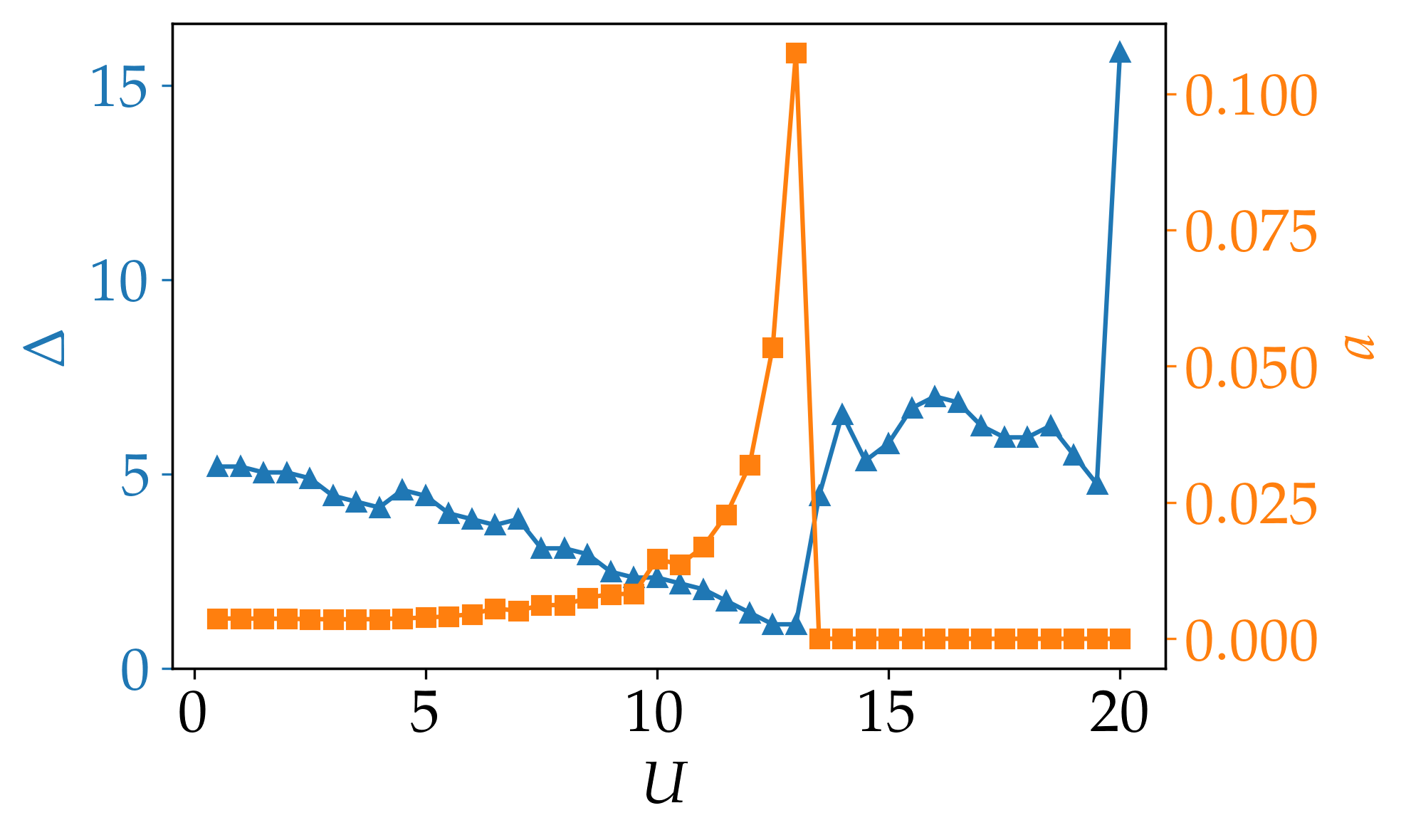}
\caption{Distance $\Delta$ between peaks of the density of states $A(\omega)$ closest to $\omega=0$ and quadratic fitting coefficient $a$ of $A(\omega)\approx a\omega^2$ at $\omega\approx0$ as a function of $U$ for $T=0.1$.}
\label{gapClosing}
\end{figure}

We conclude that in our situation, the topological Mott transition does not occur due to an emerging topologically nontrivial blind band which crosses the gap as suggested by Ref.~\onlinecite{Gurarie2011} for a possible interaction-induced topological phase transition. Rather, the topological properties stem fully from the quasiparticle bands. This requires a closing of the quasiparticle band gap. To see this quantitatively, we compute two new quantities derived from the density of states $A(\omega)$: (i) the distance between the peaks of $A(\omega)$ closest to $\omega=0$ which we denote $\Delta$. It is qualitatively equivalent to the gap of the quasiparticle bands. (ii) the coefficient of a quadratic fit of the spectral function $A(\omega)\approx a\omega^2$ at $\omega\approx0$. The coefficient $a$ is useful since it reflects the property of a semimetal that the density of states vanishes at $\omega=0$. In Fig.~\ref{gapClosing}, we show $\Delta$ as well as $a$ as a function of $U$. We indeed observe towards the expected topological phase transition point at $U\approx13$ that $\Delta$ decreases and approximately reaches zero. At the same time $a$ increases and becomes large close to $U\approx13$. Both indicates a closing of a quasiparticle gap as well as a flattening of the the semimetallic quasiparticle bands.  Ultimately, this is a possible explanation for the nonlocal annihilation of WPs: As we discussed before, the WPs do not move in the BZ while tuning the interaction strength. Instead, the topological phase transition occurs through a continuous flattening of the quasiparticle bands.

\section{Conclusion}
\label{con}
We have investigated an experimentally relevant model in the field of cold atoms in optical lattices by means of DMFT. We have calculated the double occupancy, the quasiparticle weight, as well as the density of states to determine a paramagnetic Mott insulating phase for strong Hubbard interactions. Through numerical evaluation of the Ishikawa-Matsuyama formula, which is more general than the effective topological Hamiltonian approach, we have determined the topological WSM-to-Mott-insulator transition. We investigated this topological phase transition in further detail by extracting quasiparticle  bands and blind bands which both can carry Chern numbers. It turns out that the topological phase transition occurs through a closing of the quasiparticle band gap by a continuous flattening of the semimetallic quasiparticle bands. This ultimately, enables the nonlocal annihilation of the Weyl points. The flat blind bands do not contribute to the  topological properties of the system.

\begin{acknowledgments}
The authors acknowledge enlightening discussions with Michael Pasek and Urs Gebert.
This work was supported by the Deutsche Forschungsgemeinschaft (DFG, German Research Foundation) under Project No. 277974659 via Research Unit FOR 2414. This work was also supported by the DFG via the high performance computing center LOEWE-CSC.
Tobias Grass acknowledges funding from ``la Caixa'' Foundation (ID 100010434, fellowship code LCF/BQ/PI19/11690013), ERC AdG NOQIA, Spanish Ministry MINECO and State Research Agency AEI (FIDEUA PID2019-106901GB-I00/10.13039 / 501100011033, SEVERO OCHOA No. SEV-2015-0522 and CEX2019-000910-S, FPI), European Social Fund, Fundacio Cellex, Fundacio Mir-Puig, Generalitat de Catalunya (AGAUR Grant No. 2017 SGR 1341, CERCA program, QuantumCAT U16-011424, co-funded by ERDF Operational Program of Catalonia 2014-2020), MINECO-EU QUANTERA MAQS (funded by State Research Agency (AEI) PCI2019-111828-2 / 10.13039/501100011033), EU Horizon 2020 FET-OPEN OPTOLogic (Grant No 899794), and the National Science Centre, Poland-Symfonia Grant No. 2016/20/W/ST4/00314. Jun-Hui Zheng acknowledges the support from the European Research Council via an Advanced Grant (no.\,669442 ``Insulatronics''), the Research Council of Norway through its Centres of Excellence funding scheme (project no.\,262633, ``QuSpin''),
\end{acknowledgments}

\bibliographystyle{apsrev4-1}
\bibliography{lib}

\appendix{
\section{Chern number in curvilinear coordinates}
\label{appA}
Here, we show that the analytical form of the Chern number stays invariant in an arbitrary 3d curvilinear coordinate system. We transform the expression for the Chern number which is typically defined in the cartesian BZ $(k_x,k_y,k_z)$ to the curvilinear coordinate system $(k_1,k_2,k_3)$. The flux element reads
\begin{equation}
BdS= d\bm{S}\cdot\partial_{\bm{k}}\times\bm{A}
\end{equation}
where $\bm{A}=i\langle \psi|\partial_{\bm{k}}|\psi\rangle$ is the Berry connection with $\partial_{\bm{k}}$ being the nabla operator and $|\psi\rangle\equiv|\psi(\bm{k})\rangle$ being the $\bm{k}$-dependent Bloch state. We express the Berry connection in curvilinear coordinates
\begin{equation}
\bm{A}=i\langle \psi|\partial_{k_r}|\psi\rangle\frac{\hat{\bm{e}}_r}{h_r},
\end{equation}
where $h_r=|\bm{h}_r|$ is the Lamé factor where  $\bm{h}_r=(\partial_{k_r}k_\nu)\hat{\bm{e}}_\nu$ with $\nu=x,y,z$ running over the cartesian coordinates and $r=1,2,3$ running over the curvilinear coordinates. Here, $\hat{\bm{e}}_\nu$ is the unit vector in $k_\nu$~direction and $\hat{\bm{e}}_r$ is the unit vector in $k_r$~direction. The Lamé factor is related to the metric tensor as $g_{rs}=\bm{h}_r\cdot\bm{h}_s$. 

Now, we express the curl in curvilinear coordinates
\begin{align}
\partial_{\bm{k}}\times\bm{A}=&\frac{\epsilon^{rst}}{h_sh_t}\left(\partial_{k_r}h_sA_s\right)\hat{\bm{e}}_t\\
=&\frac{i\epsilon^{rst}}{h_sh_t}\left(\partial_{k_r}\langle \psi|\partial_{k_s}|\psi\rangle\right)\hat{\bm{e}}_t
\end{align}
Here, $s,t=1,2,3$. The surface element of the surface spanned by the first and the second coordinate of the curvilinear coordinate system reads
\begin{align}
d\bm{S}&=\bm{h}_1\times\bm{h}_2dk_1dk_2\\
&=(h_1\hat{\bm{e}}_1)\times(h_2\hat{\bm{e}}_2)dk_1dk_2\\
&=h_1h_2\hat{\bm{e}}_1\times\hat{\bm{e}}_2dk_1dk_2\\
&=h_1h_2\hat{\bm{e}}_3dk_1dk_2
\end{align}
Finally, the Berry curvature element follows as
\begin{equation}
\begin{split}
d\bm{S}\cdot\partial_{\bm{k}}\times\bm{A}&=h_1h_2\hat{\bm{e}}_3\cdot\hat{\bm{e}}_t\frac{i\epsilon^{rst}}{h_rh_s}\left(\partial_{k_r}\langle \psi|\partial_{k_s}|\psi\rangle\right)dk_1dk_2\\
&=i(\partial_{k_1}\langle \psi|\partial_{k_2}|\psi\rangle-\partial_{k_2}\langle \psi|\partial_{k_1}| \psi\rangle)dk_1dk_2\\
&=-2\text{Im}\langle\partial_{k_1} \psi|\partial_{k_2}\psi\rangle dk_1dk_2
\end{split}
\label{curviChern}
\end{equation}
which has the familiar analytical form of the Chern number. Integrating the coordinates $k_1$ and $k_2$ within the respective boundaries, the resulting expression directly yields the Chern number without a complicated coordinate transformation.

\begin{figure}
\centering
\includegraphics[width=\columnwidth]{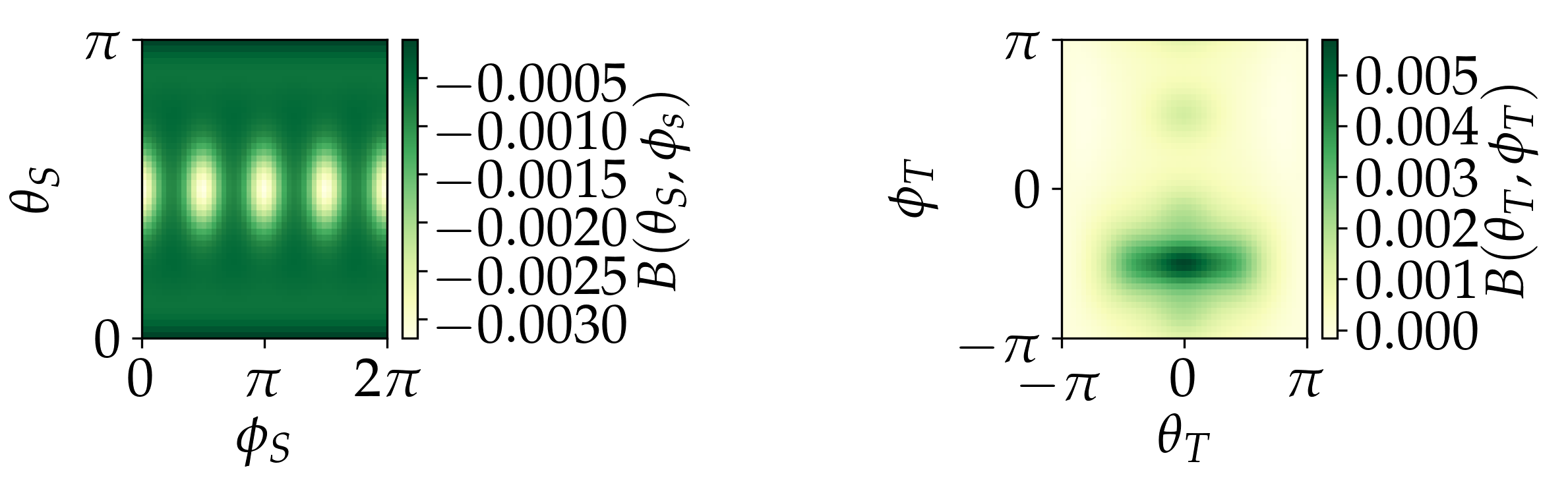}
\caption{Berry curvature on the sphere and the torus as a function of the spherical and toroidal angles.}
\label{Fig:berryCurvature}
\end{figure}

\subsection{Examples: Sphere and torus}

Let us consider two example curvilinear coordinate system to enclose the WPs, the sphere and the torus. The parametrized surfaces of the sphere $(k_1,k_2)=(\theta_S,\phi_S)$ and the torus $(k_1,k_2)=(\theta_T,\phi_T)$ can be expressed as
\begin{equation}
\begin{pmatrix}
k_x(\theta_S,\phi_S)\\
k_y(\theta_S,\phi_S)\\
k_z(\theta_S,\phi_S)\\
\end{pmatrix}
=
\bm{k}_\text{WP}+
R_S
\begin{pmatrix}
\sin(\theta_S)\cos(\phi_S)\\
\sin(\theta_S)\sin(\phi_S)\\
\cos(\theta_S)\\
\end{pmatrix}
\end{equation}
for the sphere and
\begin{equation}
\begin{split}
\begin{pmatrix}
k_x(\theta_T,\phi_T)\\
k_y(\theta_T,\phi_T)\\
k_z(\theta_T,\phi_T)\\
\end{pmatrix}
=&
\bm{k}_\text{WP}+
R_T
\begin{pmatrix}
\cos(\phi_T)\\
\sin(\phi_T)\\
0\\
\end{pmatrix}\\
&+
r_T
\begin{pmatrix}
\cos(\theta_T)\cos(\phi_T)\\
\cos(\theta_T)\sin(\phi_T)\\
\sin(\theta_T)\\
\end{pmatrix}
\end{split}
\end{equation}
for the torus. $\bm{k}_\text{WP}$ denotes the position of the WP in the BZ. The Chern number, or topological charge, then follows by substituting $(\theta_S,\phi_S)$ and $(\theta_T,\phi_T)$, respectively, for $(k_1,k_2)$ in Eq.~\eqref{curviChern}. Note that for the torus, the $\bm{k}_\text{WP}$ has to be shifted, e.g., by $R_T\hat{\bm{e}}_x$, in order to properly enclose the WP. The results for the Berry curvature as a function of $(\theta_S,\phi_S)$ and $(\theta_T,\phi_T)$, respectively, are shown in Fig.~\ref{Fig:berryCurvature} for the model in Eq.~\eqref{DubcekHam}. Integrating these Berry curvatures yields 1 and -1, respectively, according to the two different WPs enclosed. For the sphere, we have used $\bm{k}_\text{WP}=(0,\pi/2,-\pi/2)$ and $R_S=\pi/2$ and for the torus we have used $\bm{k}_\text{WP}=(-R_T,\pi/2,\pi/2)$, $R_T=\pi/6$, and $r_T=\pi/6$.

\section{Real-valuedness of the Ishikawa-Matsuyama formula}
\label{appB}
The invariant in Eq.~\eqref{IManalytical} is purely real. To show this, we  reintroduce the frequency argument and define
\begin{align}
{\Lambda}_1(i\omega_n)={G}(i\omega_n){j}_1{G}(i\omega_n){j}_2{G}(i\omega_n)(1+i\partial_{\omega_n}{\Sigma}(i\omega_n))\\
{\Lambda}_2(i\omega_n)=(1+i\partial_{\omega_n}{\Sigma}(i\omega_n)){G}(i\omega_n){j}_2{G}(i\omega_n){j}_1{G}(i\omega_n)
\end{align}
which yields
\begin{equation}
C_\text{IM}=\frac{i}{8\pi^2}\int d\bm{k}d\omega_n \text{Tr}\left[{\Lambda}_1(i\omega_n)-{\Lambda}_2(i\omega_n)\right]
\label{IMlambda}
\end{equation}
Let us consider the hermitian conjugate of ${\Lambda}_1$:
\begin{align}
&\left[{G}(i\omega_n){j}{G}(i\omega_n){j}_2{G}(i\omega_n)(1+i\partial_{\omega_n}{\Sigma}(i\omega_n))\right]^\dag\\
&=(1+i\partial_{\omega_n}{\Sigma}(i\omega_n))^\dag{G}^\dag(i\omega_n){j}_2^\dag{G}^\dag(i\omega_n){j}_1^\dag{G}^\dag(i\omega_n)\\
&=(1-i\partial_{\omega_n}{\Sigma}^\dag(i\omega_n)){G}(-i\omega_n){j}_2{G}(-i\omega_n){j}_1{G}(-i\omega_n)\\
&=(1+i\partial_{-\omega_n}{\Sigma}(-i\omega_n)){G}(-i\omega_n){j}_2{G}(-i\omega_n){j}_1{G}(-i\omega_n)\\
&={\Lambda}_2(-i\omega_n)
\end{align}
where we have used the fact that the currents ${j}_i$ are hermitian matrices as well as the symmetries of the Green's function ${G}^\dag(i\omega_n)={G}(-i\omega_n)$ and the selfenergy ${\Sigma}^*(i\omega_n)={\Sigma}(-i\omega_n)$.
We thus find that Eq.~\eqref{IMlambda} can be rewritten as
\begin{equation}
\begin{split}
C_\text{IM}&=\frac{i}{8\pi^2}\int d\bm{k}d\omega_n \text{Tr}\left[{\Lambda}_1(i\omega_n)-{\Lambda}_1^\dag(-i\omega_n)\right]\\
&=\frac{i}{8\pi^2}\int d\bm{k}d\omega_n \text{Tr}\left[{\Lambda}_1(i\omega_n)-{\Lambda}_1^\dag(i\omega_n)\right]\\
&=\frac{i}{8\pi^2}\int d\bm{k}d\omega_n \sum_l2i\text{Im}\lambda_l(i\omega_n)\\
&=-\frac{1}{4\pi^2}\int d\bm{k}d\omega_n \sum_l\text{Im}\lambda_l(i\omega_n)
\end{split}
\end{equation}
which is purely real. From the first line to the second line, we have used that we integrate over the full frequency range. In the third line we have expressed the trace of ${\Lambda}_1(i\omega_n)$ in terms of its eigenvalues $\lambda_l(i\omega_n)$.
}

\end{document}